\documentclass[12pt]{article}

\usepackage{graphicx}
\usepackage{url}

\begin{document}

\title{Public Discourse in the Web Does Not Exhibit Group Polarization}
\author{Fang Wu and Bernardo A. Huberman\\HP Laboratories\\Palo Alto, CA 94304} \maketitle

\bigskip

\begin{abstract}
We performed a massive study of the dynamics of group deliberation among
several websites containing millions of opinions on topics ranging from books
to media. Contrary to the common phenomenon of group polarization observed
offline, we measured a strong tendency towards moderate views in the course of
time. This phenomenon possibly operates through a self-selection bias whereby
previous comments and ratings elicit contrarian views that soften the previous
opinions.
\end{abstract}

\pagebreak

No aspect of the massive participation in content creation that the web enables
is more evident than in the countless number of opinions, news and product
reviews that are constantly posted on the Internet. Since these opinions play
such an important role in trust building and the creation of consensus about
many issues and products, there have been a number of recent of studies focused
on the design, evaluation and utilization of online opinion systems
\cite{CLAKR-03,dellarocas-00,GGL-06,HPZ-06} (for a survey, see
\cite{dellarocas-03}). Given the importance of group opinions to collective
social processes such as group polarization and information cascades
\cite{banerjee-92,BHW-92,chevalier-mayzlin-07,SDW-06} it is surprising that
with the exception of one study \cite{li-hitt-04}, little research has been
done on the dynamic aspects of online opinion formation. It remains unclear,
for example, whether the opinions about books, movies or societal views
fluctuate a long time before reaching a final consensus, or they undergo any
systematic changes as time goes on. Thus the need to understand how online
opinions are created and evolve in time in order to draw accurate conclusions
from that data.

Within this context we studied the dynamics of online opinion expression by
analyzing the temporal evolution of a very large set of user views, ranging
from millions of online reviews of the best selling books at \url{Amazon.com},
to thousands of movie reviews at the Internet Movie Database \url{IMDB.com}.
Surprisingly, our analysis revealed a trend that runs counter to the well known
herding effect studied under information cascades, and in the smaller instance
of group polarization. Online, a self selection mechanism is at play whereby
previous comments and ratings elicit contrarian views that soften the previous
opinions.

It is well known that in the case of group polarization, members of a
discussion group tend to advocate more extreme positions and call for riskier
courses of action than individuals who did not participate in any such
discussion \cite{asch-55,sunstein-00}. However, on the massive scale that the
web offers, we observed that later opinions in the course of time tend to show
a large difference with previous ones, thus softening the overall discourse.
This is a robust and quantitative observation for which we can only offer a
tentative explanation in terms of the cost of expressing an opinion to the
group at large.

In order to perform this study we first analyzed book ratings posted on
\texttt{Amazon.com}. Our sample consisted of the book ratings of the top 4,000
best-selling titles of \texttt{Amazon} in each of the following 12 categories,
as of July 1, 2007: arts \& photography, biographies \& memoirs, history,
literature \& fiction, mystery \& thrillers, reference, religion \&
spirituality, sports, travel, nonfiction, science, and entertainment. For each
of the 48,000 books, a series of user ratings was collected in time order,
where each rating is an integer between 1 and 5. Among the 48,000 books, 16,454
books have no less than 20 ratings, and 11,920 have an average rating above 4.

\begin{figure}
\centering
\includegraphics[width=2.5in]{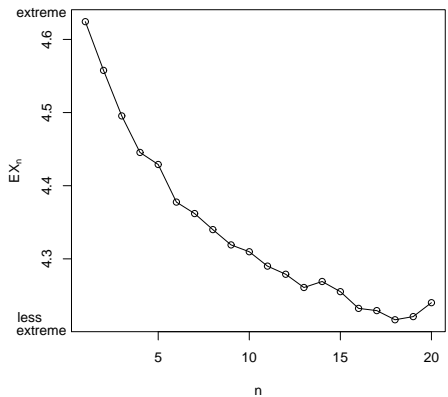}
\caption{\label{fig:amazon-mean-rating}(a) The average rating of 16,454 books
on \texttt{Amazon.com} with more than 20 reviews. $EX_n$ is the sample average
rating of all the 16,454 $n$'th ratings. As one can see from the figure, $EX_n$
decreases by 0.4 stars in 20 steps. We did not obtain enough data from low
selling books to show the opposite trend.}
\end{figure}

We first checked the average rating of the 16,454 books as a function of the
index of the rating ($n=1,\dots, 20$). As can be seen from
Fig.~\ref{fig:amazon-mean-rating}(a), $EX_n$ decreases almost linearly with
$n$, so there is a clear dynamical trend in the ratings, which corroborates the
observation reported in \cite{li-hitt-04}. Later users tend to write different
reviews from those of earlier users. Like in the experimental setup of group
polarization, an \texttt{Amazon} user observes the existing average rating of
that book before she leaves her own (usually shown at the top of the book page,
right under the title). However, as opposed to group polarization, the overall
opinion on \texttt{Amazon} tends to decrease away from the extreme ones.

\begin{figure}
\centering
\begin{minipage}{2.5in}
\centering
\includegraphics[width=2.5in]{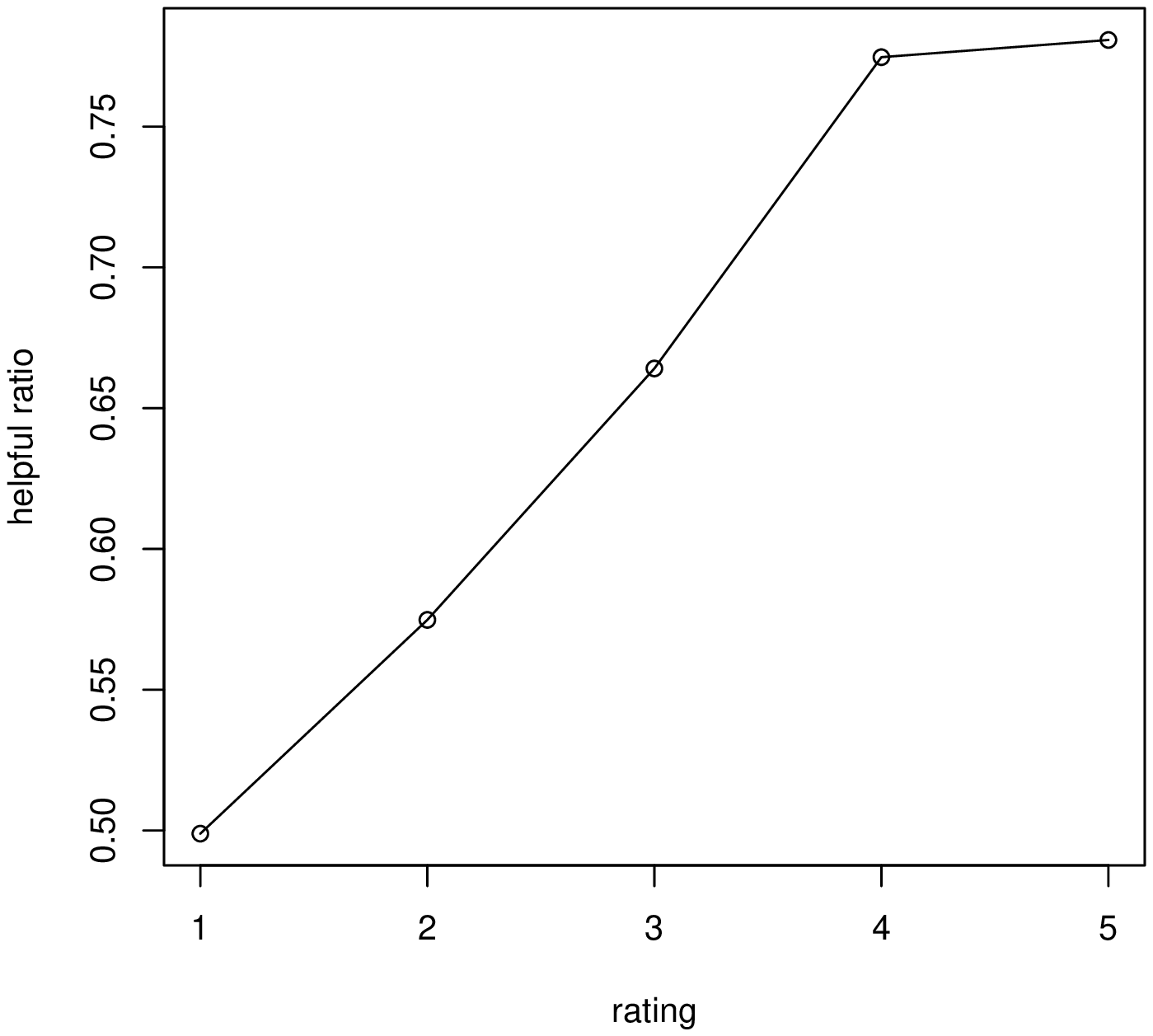}

(a)
\end{minipage}
\begin{minipage}{2.5in}
\centering
\includegraphics[width=2.5in]{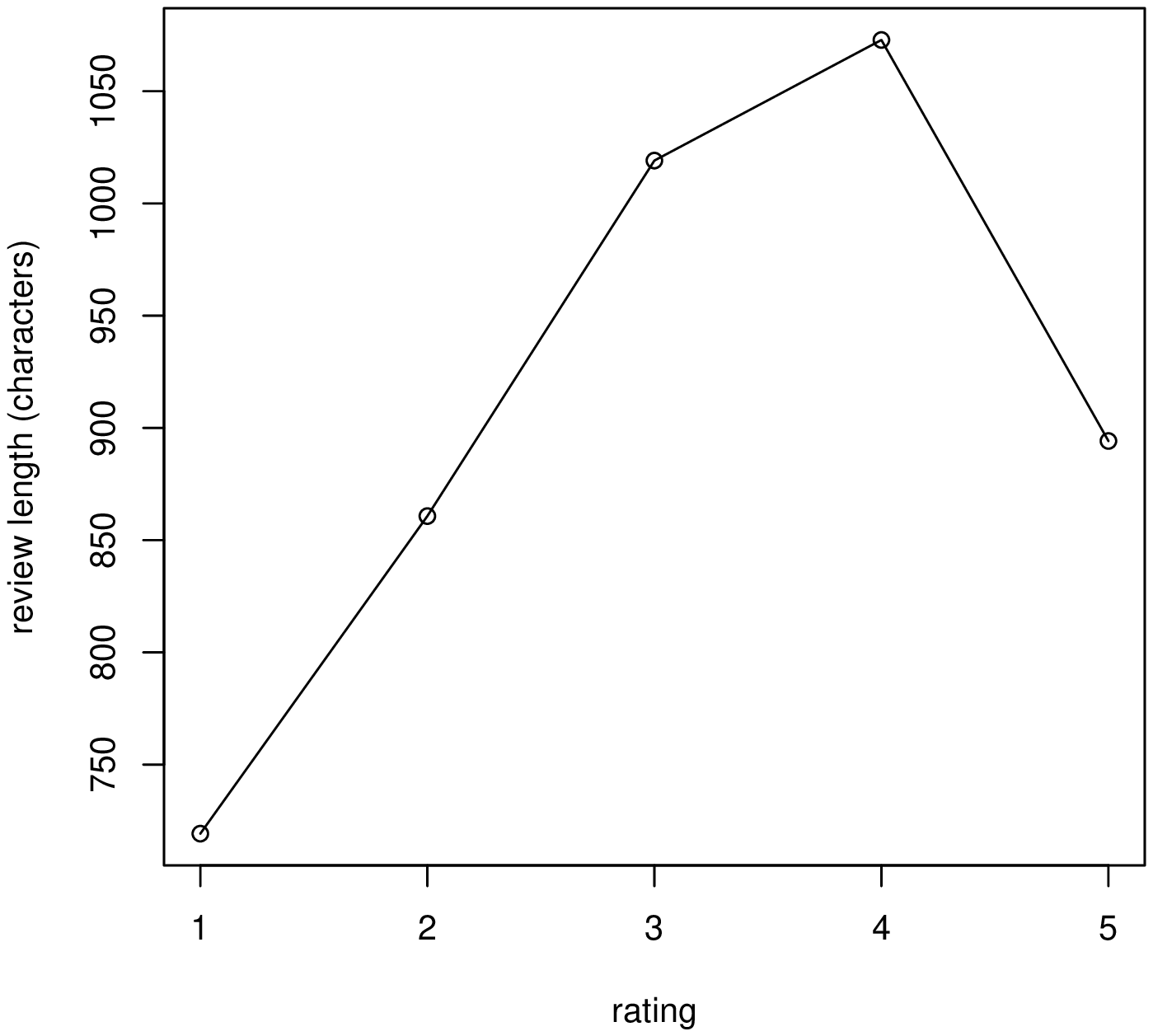}

(b)
\end{minipage}
\caption{\label{fig:amazon-review-length}(a) The average helpful ratio of five
different star ratings. (b) The average review length of five different star
ratings in the number of characters. The data is calculated for 4,000
bestselling mystery books. By comparing the two figures it is clear that people
find high ratings more helpful not just because they are longer. For instance,
5-star reviews are on average shorter than 4-star and 3-star reviews but are
nevertheless more helpful.} \end{figure}

One point to be stressed is that these results do not necessarily imply that as
time goes on the average opinion of the whole population changes, for the late
reviewers might come from a different group than the earlier ones and need not
be representative of the whole population. This is seen when plotting the
average ``helpful ratio'' as a function of star rating in
Fig.~\ref{fig:amazon-review-length} for users of \texttt{Amazon}. As can be
seen, the whole population finds high ratings in general more helpful than low
ratings, implying that the majority of the population does not necessarily
agree with the low ratings. This additional data suggests that rather than
indicating a real opinion shift in the whole population, the observed dynamic
trend is more of an expression bias.

On reflection, it is rather surprising that people contribute opinions and
reviews of topics which have already been extensively covered by others. While
posting views is easy to understand when it involves no effort, like clicking
on a button of a website, it is more puzzling in situations where it is costly,
such as composing a review.\footnote{When a user of \texttt{Amazon} decides to
review a book, she is required to write a short paragraph of review in addition
to a simple star rating. The average word count of \texttt{Amazon} reviews is
181.5 words \cite{ketzan-02}, so the cost of opinion expression is indeed
high.} If the opportunity to affect the overall opinion or rating diminishes
with the number of published ones, why does anyone bother to incur the cost of
contributing yet another review? From a rational choice theory point of view,
if the utility to be gained does not outweigh the cost, people would refrain
from expressing their views. And yet they do. This is reminiscent of the well
analyzed voter's paradox \cite{downs-57,riker-ordeshook-68,schuessler-00},
where a rational calculation of their success probability at determining the
outcome of an election would make people stay home rather than vote, and yet
they show up at the polls with high turnout rates. In contrast to a political
election, there is no concept of winning in online opinion systems. Rather, by
contributing her own opinion to an existing opinion pool, a person affects the
average or the distribution of opinions by a marginal amount that diminishes
with the size of that pool.

One possible explanation for these results is that in cases like \url{Amazon},
people will derive more utility the more they can influence the overall rating,
as in the voter's paradox. To be precise, in cases where users' opinions can be
quantified and aggregated into an average value, the influence of an online
opinion can be measured by how much its expression will change the average
opinion. Suppose that $n$ users have expressed their opinions, $X_1, \dots,
X_n$, on a given topic at a website, with $X_i$ denoting the quantified value
of the $i$'th opinion. If the $(n+1)$'th person expresses a new opinion
$X_{n+1}$, it will move the average rating to \begin{equation} \bar X_{n+1} =
\frac{n\bar X_n + X_{n+1}}{n+1}, \end{equation} and the absolute change in the
average rating is given by \begin{equation} \label{eq:average-diff} |\bar
X_{n+1}- \bar X_n| = \frac{|X_{n+1}-\bar X_n|}{n+1}.
\end{equation}
Thus a person is more likely to express her opinion whenever $|X_{n+1}-\bar
X_n|$ is large --- an opinion is likely to be expressed if it deviates by a
significant amount from those already stated. Indeed, what is the point of
leaving another 5-star review after one hundred people have already done
so?\footnote{This point has also been made within the ``brag-and-moan'' model
\cite{dellarocas-narayan-06,HPZ-06} which assumes that consumers only choose to
write reviews when they are very satisfied with the products they purchased
(brag), or very disgruntled (moan). Note however, that the brag-and-moan model
is static and thus predicts that $\bar X_n$ is constant over time, in
contradiction with the observed dynamical trends.}

\begin{figure}
\centering
\includegraphics[width=2.5in]{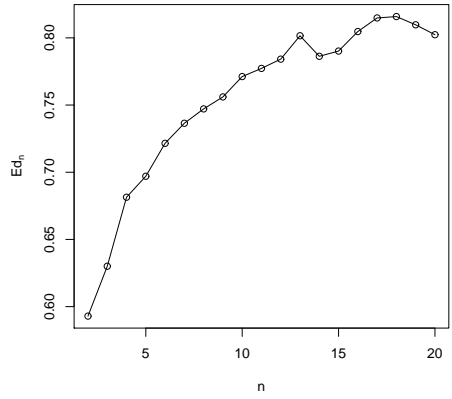}
\caption{\label{fig:amazon-deviation}The average deviation of \texttt{Amazon}
ratings increases with the number of people.}
\end{figure}

In order to test this hypothesis, we measured directly how much one's rating
deviates from the observed average rating. We plot the expected
\emph{deviation} $Ed_n = E|X_n - \bar X_{n-1}|$ as a function of $n$ in
Fig.~\ref{fig:amazon-deviation}, where $X_n$ is the rating left by the $n$'th
user, and $\bar X_{n-1}$ is the average rating the $n$'th user observes. As can
be seen, $Ed_n$ increases with $n$. Since the expected deviation $Ed_n$ of an
i.i.d.~sequence normally \emph{decreases} with $n$, this increasing trend is
indeed significant. This again supports our conjecture that those users who
disagree from the public opinion will be more willing to express themselves and
thus soften the overall opinion of a given book.

\begin{figure}
\begin{center}
\includegraphics[width=3in]{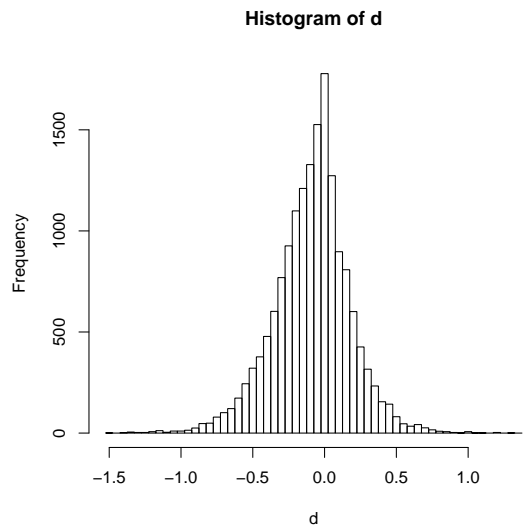}
\caption{\label{fig:amazon-hist}Histogram of the change in average book ratings
($d=\bar X_{20}-\bar X_{10}$) on \texttt{Amazon.com}. Most of the changes are
negative, testifying a declining trend in the average ratings.} \end{center}
\end{figure}

Next we examined whether this dynamical trend is still prominent at the level
of each individual book. We defined $d=\bar X_{20}-\bar X_{10}$ as a measure of
the change in a book's rating over time. The histogram of 16,454 $d$'s is shown
in Fig.~\ref{fig:amazon-hist}. As can be seen, most of the changes are
negative. A $t$-test of the alternative hypothesis ``$d<0$'' yields a $p$-value
less than $0.0001$, which further confirms the declining trend.

While our hypothesis of a costly expression bias seems to explain the softening
of opinions observed in \texttt{Amazon}, it would be more conclusive if one
could conduct a test that directly compares people's opinions expressed at
different cost levels. In order to address this issue we conducted a study of
\texttt{IMDB.com} (The Internet Movie Database). Unlike users of
\texttt{Amazon} who are required to write a review when rating a book, users of
\texttt{IMDB} are free to \emph{choose} the effort level when reviewing a
movie. Specifically, after observing the current average rating of a movie, a
user can either submit a quick rating by clicking on a scale of 10 stars, or
can make the extra effort involved in writing a comment between 10 and 1000
words.

Our study focused on two sets of movie titles. The first consists of the 50
most top-rated movies released after year 2000, which we call the ``good
movies'', and the second consists of the 50 most low-rated, which we call the
``bad movies''. For each movie we know its average rating (taken among all
ratings with or without a comment), as well as the value and date-stamp of its
each commented rating, but we do not have any specific information about each
uncommented rating.

\begin{figure}
\centering
\begin{minipage}{2.5in}
\centering
\includegraphics[width=2.5in]{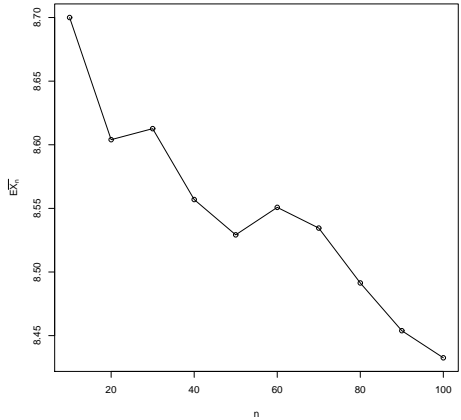}

(a) Good movies
\end{minipage}
\begin{minipage}{2.5in}
\centering
\includegraphics[width=2.5in]{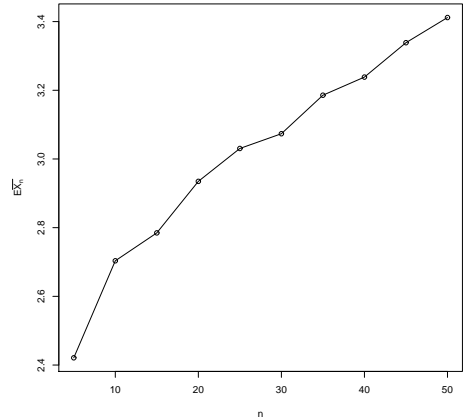}

(b) Bad movies
\end{minipage}
\caption{\label{fig:movie-ratings}Average rating associated with a comment of
the (a) good and (b) bad movies, as a function of the number of existing
ratings. It can be seen that good movies tend to receive lower ratings as time
goes on, and bad movies tend to receive higher ratings.} \end{figure}

\begin{figure}
\centering
\begin{minipage}{2.5in}
\centering
\includegraphics[width=2.5in]{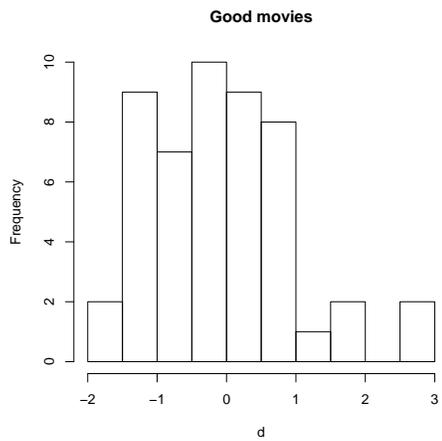}

(a)
\end{minipage}
\begin{minipage}{2.5in}
\centering
\includegraphics[width=2.5in]{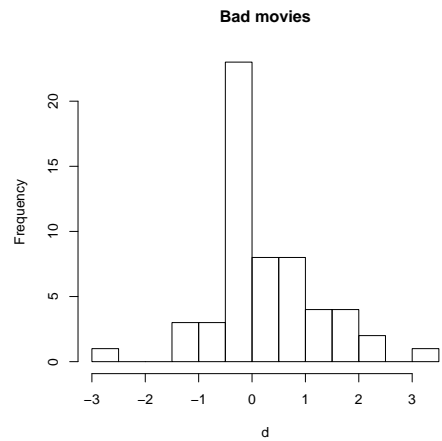}

(b)
\end{minipage}
\caption{\label{fig:idmb-hist}Histogram of $d=\bar X_{10}-\bar X_5$ for the
good movies and bad movies.} \end{figure}

The trend of the ratings associated with comments of the two sets of movies is
shown in Fig.~\ref{fig:movie-ratings}. Similar to \texttt{Amazon}, a softening
of the expressed view is once again observed for both sets. Two histograms of
$d=\bar X_{10}-\bar X_5$ for the good movies and the bad movies are shown in
Fig.~\ref{fig:idmb-hist}. A $t$-test of the alternative hypothesis $d<0$ for
the good movies yields a $p$-value 0.44. A $t$-test of $d>0$ for the bad movies
yields a $p$-value 0.018. While it is not too reliable to conclude that good
movies tend to receive lower ratings over time, it is safer to conclude that
bad movies accumulate higher ratings as time goes on.

\begin{figure}
\centering\includegraphics[width=3in]{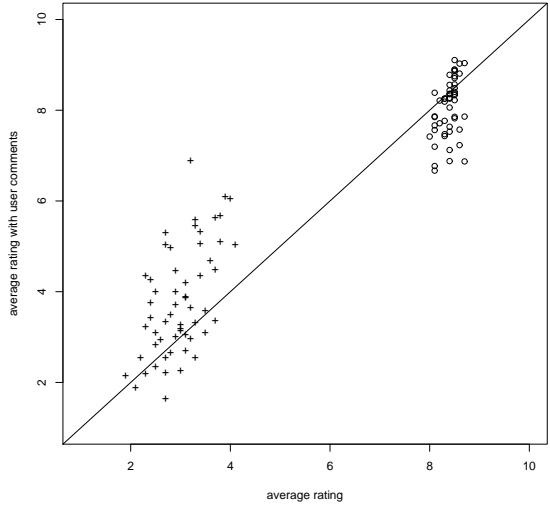}
\caption{\label{fig:commented-ratings}Expression bias of commented ratings.
Each point in this figure corresponds to one movie title. The horizontal
coordinate represents the movie's overall average rating ($\bar r$) taken over
both commented and uncommented ratings. The vertical coordinate represents the
movie's average rating taken over only commented ratings ($\bar r_c$). Good and
bad movies are represented by circles and crosses, respectively. Clearly, those
users who spend the additional cost to write a comment tend to speak oppositely
to the majority. A $t$-test of the alternative hypothesis that $\bar r_c<\bar
r$ for good movies and a similar $t$-test of $\bar r_c>\bar r$ for bad movies
both yield a $p$-value less than $0.001$.} \end{figure}

We also examined the difference between the overall average rating (with or
without a comment) and the average rating associated with a comment for each
movie, and the result is shown in Fig.~\ref{fig:commented-ratings}. It can be
seen that those who decide to spend the time to write a comment tend to speak
differently from the majority users, who simply leave a star rating without any
justification. Fig.~\ref{fig:commented-ratings} is thus a direct verification
of our hypothesis that high cost induces expression bias.

\medskip
These results show that in the process of articulating and expressing their
views online, people tend to follow a different pattern from that observed in
information cascades or group polarization. What is observed is an anti
polarization effect, whereby previous comments and ratings elicit contrarian
views that soften the previous opinions. This is in contrast to the phenomenon
of herding and opinion polarization observed in both group dynamics and online
sites.\footnote{We point out that in a website like \texttt{Jyte.com}, where it
takes only one click to agree or disagree with an arbitrary claim, we did see a
strong group polarization \cite{SDW-06}. It is possible that the latter is due
to the fact that such a vote is costless compared to the opinions on
\texttt{Amazon} and \texttt{IMDB}.}

In closing, besides their intrinsic novelty, these results throw a cautionary
note on the interpretation of online public opinion. This is because a simple
change in the order or frequency of given sets of views can change the ongoing
expression in the community, and thus the perceived collective wisdom that new
users will find when accessing that information.

\bibliographystyle{plain}
\bibliography{rating}

\end{document}